\documentclass[twocolumn,showpacs,prl,superscriptaddress]{revtex4}
\usepackage{amsmath}
\usepackage{mathrsfs}
\usepackage{graphicx}

\begin{document}
\draft

\title{Charge Detection in a Closed-Loop Aharonov-Bohm Interferometer}%

\author{Gyong Luck Khym and Kicheon Kang}\email{kckang@chonnam.ac.kr}
\affiliation{Department of Physics and Institute for Condensed
Matter Theory, Chonnam National University, Gwangju 500-757,
Korea}

\date{\today}

\begin{abstract}
We report on a study of complementarity in a two-terminal
$closed$-$loop$ Aharonov-Bohm interferometer. In this
interferometer, the simple picture of two-path interference cannot
be applied. We introduce a nearby quantum point contact
to detect the electron in a quantum dot inserted in the
interferometer. We found that charge detection reduces but does not
completely suppress the interference even in the limit of perfect
detection. We attribute this phenomenon to the unique nature of the
closed-loop interferometer. That is, the closed-loop interferometer
cannot be simply regarded as a two-path interferometer because of
multiple reflections of electrons. As a result, there exist
indistinguishable paths of the electron in the interferometer and
the interference survives even in the limit of perfect charge
detection. This implies that charge detection is not equivalent to
path detection in a closed-loop interferometer. We also discuss the
phase rigidity of the transmission probability for a two-terminal
conductor in the presence of a detector.
\end{abstract}


\maketitle

Complementarity in quantum theory is well described in a two-path
interferometer such as Young's double slit interferometer. In a
two-path interferometer with a `which-path' detector, observation of
the interference pattern and the acquisition of which-path
information are mutually
exclusive~\cite{feynman65,stern90,scully91}. Most of the work on
understanding this kind of interferometer has been carried out in
optical systems with photons~\cite{zeilinger99}.
Only recently has it become possible to investigate the
complementarity of electrons in solid-state
circuits~\cite{buks98,sprinzak00,kalish04}. The interference is
shown by the oscillation of conductance as a function of magnetic
flux in an Aharonov-Bohm (AB) interferometer with a quantum dot (QD)
inserted in one of its arms. This AB oscillation of conductance has
been observed both in a closed~\cite{yacoby95} and in an
open-geometry~\cite{schuster97}. The open-geometry AB interferometer
of Ref.~\onlinecite{schuster97} can be regarded as a solid-state
version of Young's double slit interferometer. A mesoscopic
which-path interferometer has been demonstrated by using an
open-geometry AB interferometer containing a quantum dot (QD) with a
nearby quantum point contact (QPC) used as a which-path
detector~\cite{buks98}. The QPC interacts with the QD and is able to
detect a single charge in the QD. The detection is made through the
QD-charge dependence of the scattering coefficients at the QPC. This
results in decoherence of the charge state of the QD and suppression
of the AB oscillation. The suppression strength is controlled
through the voltage across the QPC. Different setups for the
controlled dephasing experiment have been also demonstrated by using
QD-QPC hybrid structures~\cite{sprinzak00,kalish04}.

It is obvious that charge detection is equivalent to the path detection
in a double-slit (or two-path) AB
 interferometer investigated in Ref.~\onlinecite{buks98}. On the
 other hand, it is an interesting question as to what would happen in a
 closed-loop AB interferometer of the type studied in Ref.~\onlinecite{yacoby95}
 when a charge detector is attached. In a
 closed-loop AB interferometer, the conductance through the system
 is not simply given by interference between electron
 transmission through the two direct paths~\cite{yeyati95}.
 Therefore, a charge detection may not be equivalent to the path
 detection in this closed-loop interferometer.

\begin{figure}[b]
\includegraphics[width=75mm]{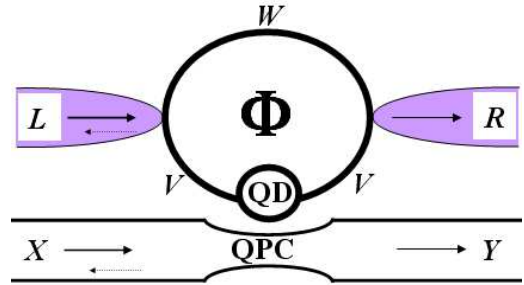} \caption{\label{schematic}
Schematic figure of the model system: A closed-loop Aharonov-Bohm
interferometer with a quantum point contact detector. A quantum
dot is inserted in the interferometer and coupled to the quantum
point contact which enables detection of the charge state in the
quantum dot.}
\end{figure}
In this Letter, we report on our investigation of complementarity
in a closed-loop AB interferometer, where the simple picture of
Young's double slit is invalid. A QD is embedded in one arm of the
interferometer, and is coupled to a QPC being used as a charge
detector (See Fig.~\ref{schematic}). In contrast to that of a
two-path interferometer, we show that the AB oscillation is not
completely suppressed even in the limit of perfect charge
detection. This feature originates from the fact that charge
detection does not entirely determine the path of an electron in a
closed-loop interferometer.

The Hamiltonian of a closed-loop AB interferometer with a QD inserted in
one of its arms is given by $ H_{AB} = H_{0} + H_{1}$, where
\begin{subequations}
\begin{eqnarray}
 H_{0} &=& \epsilon_{d} d^{\dagger}d + \sum_{\alpha} \sum_{k} \epsilon_{k} c^{\dagger}_{\alpha k}c_{\alpha k}
     \;, \label{H0}  \\
 H_{1} &=& \sum_{k} [( W c^{\dagger}_{Lk}c_{Rk} + Vd^{\dagger}c_{Lk}
 + Vc^{\dagger}_{Rk}d   ) + h.c. ]\;.~~~\label{H1}
\end{eqnarray}
$H_{0}$ denotes a QD and the two leads ($L,R$). $H_{1}$
describes transfer of electrons between the subsystems. The operator
$d$ ($d^{\dagger}$) annihilates (creates) an electron in the QD with
energy $\epsilon_d$. The operator $c^{\dagger}_{\alpha k}$ and
$c_{\alpha k}$ ($\alpha= L, R$) refer to states in the lead
$\alpha$. The hoping amplitudes $V$ and $W$ are chosen as $V=|V|$
and $W=|W|e^{i\varphi}$ with $\varphi$ being the AB phase.
\end{subequations}

A nearby QPC detector close to the QD is introduced. Because of
the Coulomb interactions between the QD and the QPC, the electron
state in the QPC depends on the trajectory of electron in the
interferometer. Accordingly, dephasing of the QD electron state
takes place. Various theoretical approaches have been reported
that address this
issue~\cite{aleiner97,silva01,levinson97,Gurvitz97,hacken98,levinson00,Stodolsky,Butt-Martin,Korotkov-Averin,Pilgram,kang05}.
In a two-path which-path interferometer, the visibility reduction
factor of the AB oscillation is proportional to the dephasing
rate. Naturally, AB oscillation in a two-path interferometer
disappears in the perfect detection limit in which the two
detector states are orthogonal.

\begin{figure}[b]
\includegraphics[width=75mm]{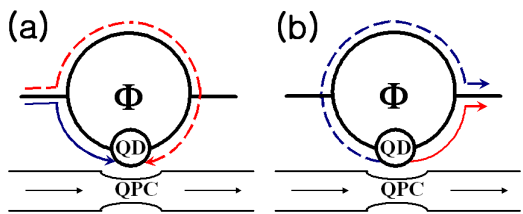}
\caption{\label{indistinguishable} Examples of indistinguishable
paths for an electron transport with charge detection at the QD.
(a) Two indistinguishable paths from the left lead to the QD. (b)
Two indistinguishable paths from the QD to the right lead.}
\end{figure}
The situation is different for a closed-loop interferometer. Quantum
interference originates from the indistinguishability of two or more
events. In fact, there exist {\em indistinguishable paths} of the
electron in the closed-loop interferometer despite perfect charge
detection. This is because charge detection does not provide
complete information on the electron's path. Examples are shown in
Fig.~2. Considering an electron injected from the left lead which is
detected by the QPC at the QD (Fig.~2(a)), one cannot determine
whether this electron came from the left (full arrow) or from the
right (dashed arrow) arm. Similarly, for an electron detected at the
QD and then absorbed in the right lead, one cannot determine a
definite path as shown in Fig.~2(b). This kind of
indistinguishability does not exist in a double-slit type
interferometer where a charge detection is equivalent to path
detection. In our setup, these kind of events contribute to the AB
oscillation of the conductance through the interferometer in the
perfect charge detection limit.

Let us now describe quantitatively the hybrid system of a
closed-loop interferometer and a QPC detector. For simplicity, we
assume that the two electrons, one from lead $L$ of the
interferometer and the other from lead $X$ of the detector (See
Fig.~\ref{schematic}), are simultaneously injected. Also, our
discussion is restricted to the off-resonance limit of the QD. Then,
upon a scattering of the two electrons, we can write the
two-particle state as
\begin{equation}
 |\psi \rangle \simeq |\phi_0\rangle_e \otimes | \chi_0 \rangle_d +
                 |\phi_1 \rangle_e \otimes | \chi_1 \rangle_d \;,
\label{eq:psi}
 \end{equation}
where $|\phi_0\rangle_e$ denotes all possible paths which do not go
into the QD. The state $|\phi_1 \rangle_e$ includes all processes
that includes the leading (second) order tunneling through the QD.
Higher order tunneling processes can be neglected in the
off-resonance limit. Note that $|\phi_0 \rangle_e$ and $|\phi_1
\rangle_e$ include multiple reflections at the contacts between the
leads and the interferometer. These multiple reflections make the
system different from a two-path interferometer. $|\chi_0 \rangle_d$
and $|\chi_1 \rangle_d$ represent the corresponding detector states.
These states can be written as
\begin{equation}
 |\chi_i \rangle_d = \bar{r}_i|X\rangle + \bar{t}_i|Y\rangle \;\;
  (i\in0,1),
\end{equation}
where $\bar{r}_i$ and $\bar{t}_i$ are the $i$-dependent reflection
and transmission amplitudes, respectively. $|X\rangle$ and
$|Y\rangle$ are the states of the electron being at lead $X$ and
$Y$, respectively.


For the state $|\psi\rangle$ of Eq.(\ref{eq:psi}), the probability
of finding an electron at lead $R$ (equivalent to the transmission
probability) is given as
\begin{equation}
T_{LR} = \int {\Big(} - \frac{\partial f }{ \partial \epsilon}
{\Big )} \langle \psi |R\rangle \langle R| \otimes I_d
|\psi\rangle  d\epsilon \;,
\end{equation}
where $f$ is the Fermi distribution function. $|R\rangle$
corresponds to the state of the electron being at lead $R$. $I_d$ is
the identity operator that acts only on the detector. At zero
temperature, one finds that
\begin{subequations}
\label{eqs:TLR}
\begin{equation}
 T_{LR} = |t_0|^2 + |t_1|^2
 + 2{\rm Re}[{\lambda t_0^{*} t_1}]
 \;, \label{TLR}
\end{equation}
where $t_0 = \langle R | \phi_0\rangle_{e}$ and
$t_1 = \langle R | \phi_1 \rangle_{e}$ are the transmission amplitudes
for the state $|\phi_0\rangle_e$ and $|\phi_1\rangle_e$, respectively.
The constant $\lambda$ given as
\begin{equation}
 \lambda = {_d\langle\chi_0|\chi_1\rangle}_d =
 \overline{r}_0^{*}\overline{r}_1 + \overline{t}_0^{*}\overline{t}_1
\end{equation}
is a measure of the indistinguishability between the states,
$|\phi_0\rangle_e$ and $|\phi_1\rangle_e$.

\begin{figure}[b]
\includegraphics[width=75mm]{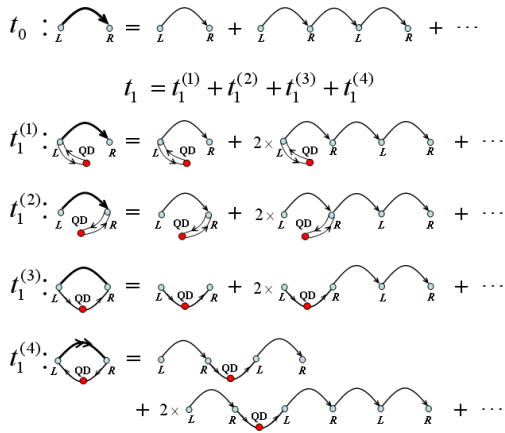} \caption{\label{diagram} Diagrams
which represents the partial waves of an electron moving from lead
$L$ to $R$. $t_0$ represents the electron transition through the
free arm. $t_1$ includes all possible diagrams with a single
tunneling event through the QD. These can be classified into four
different contributions, namely $t_1^{(1)}$, $t_1^{(2)}$,
$t_1^{(3)}$, and $t_1^{(4)}$ as shown in the figure.}
\end{figure}

In general, transmission amplitude, $t_i$ ($i\in 0,1$), (in the
absence of a `detector') can be obtained from Green's function $G_i$
($i\in0,1$) using the relation~\cite{fisher81},
\begin{equation}
 t_i = i\hbar\nu G_i \;,
\end{equation}
where $\nu$ is the Fermi velocity. In our case, the Green's
functions, $G_i$, (and the transmission coefficients $t_i$) which
correspond to the states $|\phi_i\rangle_e$ are calculated by using
perturbation expansion on the hopping part of the Hamiltonian ($H_1$).
Diagrams for the infinite series of this perturbation expansion are
given in Fig.~3. After some algebra, one finds that
\begin{equation}
t_0 = \frac{-2ix}{1+ x^2}e^{-i\varphi} \;,
\end{equation}
where $x= \pi\rho |W|$ with $\rho$ being the density of states at
the Fermi energy,
and
\begin{equation}
 t_1 = \frac{\Gamma}{\epsilon_d}t_0
   \left( 2i-\frac{e^{i\varphi}}{x} + xe^{-i\varphi} \right) \;,
\end{equation}
\end{subequations}
where $\Gamma$ is the effective resonance width of the QD level given as
$\Gamma=\pi\rho|V|^2/(1+x^2)$.

\begin{figure}[b]
\includegraphics[width=75mm]{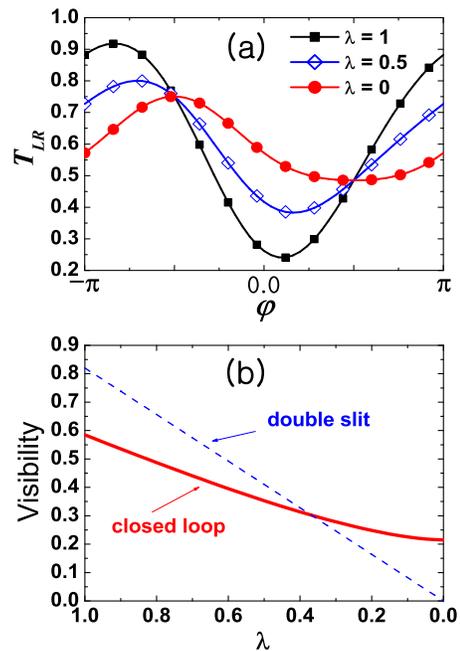} \caption{\label{re0p07}
(a) The transmission probability through the closed-loop
interferometer as a function of the AB Phase, and (b) the
visibility of interference pattern (full line) as a function of
$\lambda$, for $x =0.4$, $V = 0.75|W|$, $\epsilon_d = 1.25|W|$,
and $\Gamma/\epsilon_d = 0.155 $.  For comparison, the visibility
of `double-slit' interferometer (dashed line) is also shown for
the same parameters.}
\end{figure}

Fig.~\ref{re0p07}(a) shows the AB oscillations of the transmission
probability. As one can see, the amplitude of the AB oscillation is
reduced as $\lambda$ decreases but does not vanish even in the limit
of $\lambda=0$. The visibility ${\cal V}$ defined as ${\cal V} =
(\max{(T_{LR})}-\min{(T_{LR})})/(\max{(T_{LR})}+\min{(T_{LR})})$ is
shown in Fig.~\ref{re0p07}(b) as a function of $\lambda$. The
visibility is, in general, reduced as $|\lambda|$ decreases.
However, the oscillation is not entirely suppressed even in the
limit of perfect charge detection ($\lambda=0$), in contrast to that
of a `double-slit' type interferometer. Note that the remaining
visibility in the $\lambda\rightarrow0$ limit comes from the fact
that $|t_1|^2$ depends on the AB phase $\varphi$. In fact, the AB
interference of $|t_1|^2$ results from the indistinguishability of
various paths of the electron as shown in Fig.~3. This is the origin
of the finite visibility for $\lambda=0$ where the interference
between the states $|\phi_0\rangle_e$ and $|\phi_1\rangle_e$
vanishes.

One might interpret the visibility reduction through charge
detection in the following way: Coulomb interaction between the
electrons in the QD and in the detector disturbs the motion of an
electron in the QD. This disturbing then results in an uncertainty
of the electron's phase. This `random phase' washes out the
interference. In this interpretation, however, the closed-loop
interferometer would not be different from the two-path
interferometer. That is, any electron would loose its phase
coherence whenever it passes through the QD, and the visibility of
the AB oscillation vanishes in the strong detection limit. Then, the
closed-loop interferometer would not show any essential difference
from the double-slit interferometer, in spite of nontrivial
electronic paths. Our results (Eq.~\ref{eqs:TLR} and Fig.~4) clearly
indicate that the `random-phase' interpretation does not apply to
the charge detection in a closed-loop interferometer.

{\em About the phase rigidity:} It is well known that the
transmission probability for a two-terminal conductor (without a
detector) should satisfy the relation
$T_{LR}(-\varphi)=T_{LR}(\varphi)$, so that a `phase rigidity'
exists~~\cite{yeyati95}. This phase rigidity is not satisfied in our
result as one can see in Fig.~\ref{re0p07} (a). One reason for the
breaking of the phase rigidity is because of our approximations that
neglect higher order tunneling processes. These approximations do
not fully take into account the unitarity of electron scattering.
Therefore, it naturally breaks the phase rigidity. However, there is
a more fundamental reason for the phase rigidity breaking. That is,
the phase rigidity does not exist in a system interacting with a
detector. In the following, we develop a general argument on this
lack of phase rigidity.

 Let us consider two particle injection: one from lead $L$ of the
interferometer, and the other from lead $X$ of the detector. The two
electrons interact with each other at the QD-QPC contact region.
Upon a scattering, the two electrons have four possible
configurations, namely
 $| 1 \rangle \equiv |L \rangle \otimes |X \rangle$, $|2 \rangle
 \equiv |L \rangle \otimes |Y \rangle$,
 $|3 \rangle \equiv |R \rangle \otimes |X \rangle$,
 and $|4 \rangle \equiv | R \rangle \otimes |Y \rangle$.
 With this representation, the initial state can be written as
 $|\psi_{in}\rangle=|1\rangle$. After scattering, the two-particle
state is given by
\begin{equation}
 |\psi(\varphi)\rangle = \hat{S}(\varphi)|\psi_{in}\rangle
  = \sum_{i=1}^4 S_{i1}(\varphi)|i\rangle\,,
\end{equation}
where $\hat{S}$ denotes the two-particle scattering matrix and
$S_{ij}\equiv \langle i|\hat{S}|j\rangle$. Note that $\hat{S}$
cannot be written as a direct product of two single-particle
scattering matrices because of the interaction between the two
subsystems. The reciprocity relation of $\hat{S}$ gives the
constraint $S_{ij}(\varphi) = S_{ji}(-\varphi)$. Also, the unitarity
of $\hat{S}$ requires $\sum_{i=1}^{4} |S_{i1}(\varphi)|^2 =
\sum_{i=1}^{4} |S_{1i}(\varphi)|^2 =1$. The transmission probability
in the interferometer is given by
\begin{equation}
 T_{LR}(\varphi) =
   \langle\psi(\varphi)|R\rangle\langle R|\otimes I_d|\psi(\varphi)\rangle
   = |S_{31}(\varphi)|^2 + |S_{41}(\varphi)|^2\;.
 \label{eq:TLR2}
\end{equation}
From Eq.(\ref{eq:TLR2}) together with the unitarity and the
reciprocity relations, we find that the phase rigidity
$T_{LR}(-\varphi)=T_{LR}(\varphi)$ is satisfied if
$|S_{21}(\varphi)|^2 =|S_{12}(\varphi)|^2$. However, this condition
is not valid in general in the presence of two-particle
interactions. (That is, $S_{21}=S_{12}$ only when the two systems
are independent.) Therefore, we conclude that {\em the phase
rigidity of a two-terminal conductor is not enforced, in general, if
the conductor is interacting with another system}.

In conclusion, we studied the influence of charge detection on the
conductance oscillation in a two-terminal $closed$-$loop$ AB
interferometer where the simple picture of a double-slit
interference cannot be applied. We found that charge detection
reduces but does not fully suppress the interference even in the
limit of perfect detection. This interesting property originates
from multiple reflections of an electron in the two-terminal
interferometer. Because of the multiple reflections of an electron
in the interferometer, full information about the electronic path
cannot be obtained through charge detection. This
$indistinguishability$ of electronic paths results in the AB
oscillation. Furthermore, based on a general argument, we pointed
out that the phase rigidity is not enforced in a two-terminal
conductor if the conductor is interacting with another subsystem.

\acknowledgements%
We acknowledge helpful discussions and comments from Y.-C.~Chung and
H.-W.~Lee. This work was supported by the Korea Research Foundation
(KRF-2004-202-C00166, KRF-2005-070-C00055).


\end{document}